\documentclass[10pt,twocolumn,english,aps,manuscript]{revtex4}
\usepackage[T1]{fontenc}
\usepackage[utf8]{inputenc}
\usepackage{amsthm}
\usepackage{amsmath}
\usepackage{graphicx}
\usepackage{amssymb}

\theoremstyle{plain}

\usepackage{graphics}

\usepackage{indentfirst}

\usepackage{enumerate}

\begin{document}

\title{Properties of entangled photon pairs generated in periodically poled nonlinear
crystals}

\author{Ji\v{r}\'{\i} Svozil\'{\i}k  and Jan Pe\v{r}ina Jr.}
\affiliation{Joint Laboratory of Optics of Palack\'{y} University
and Institute of Physics of Academy of Sciences of the Czech
Republic, 17. listopadu 50A, 772 07 Olomouc, Czech Republic}

\begin{abstract}
Using a rigorous quantum model a comprehensive study of physical
properties of entangled photon pairs generated in spontaneous
parametric down-conversion in chirped periodically-poled crystals
is presented. Spectral, temporal, as well as spatial
characteristics of photon pairs are analyzed. Spectral bandwidths,
photon-pair flux, and entanglement area can be effectively
controlled by chirping. Quantification of entanglement between
photons in a pair is given. Splitting of entanglement area in the
transverse plane accompanied by spectral splitting has been
revealed. Using the model temperature dependencies of the
experimental intensity profiles reported in literature have been
explained.
\end{abstract}


\maketitle

\section{Introduction}

The beginning of nonlinear quantum optics can be dated back to the
sixties of the 20th century, when the first nonlinear optical
phenomena were observed \cite{PhysRevLett.7.118} owing to the
discovery of the laser. The next milestone was reached twenty
years later by Hong, Ou and Mandel who observed strong quantum
correlations between photons in a photon pair generated in the
process of spontaneous parametric frequency down-conversion
\cite{PhysRevA.34.3962}. These photon pairs represent the
elementary quantum objects that can be correlated in time, space,
momentum, energy, polarization or even orbital angular momentum.
They have been used in many applications including tests of Bell's
inequalities \cite{PhysRevLett.81.5039}, quantum cryptography
\cite{PhysRevLett.68.557}, quantum teleportation
\cite{PhysRevLett.80.1121,Nature390}, and metrology
\cite{1999PhT....52a..41M}.

Many different structures emitting entangled photon pairs have
been developed. Nonlinear bulk crystals \cite{PhysRevA.50.5122}
and nonlinear wave-guides belong to standard sources of these
photon pairs nowadays. Also new sources based on photonic-crystal
fibers \cite{Li2005,Fulconis2005}, layered photonic structures
\cite{PerinaJr2006,Centini2005,Vamivakas2004,PerinaJr2007},
nonlinear cavities \cite{Shapiro2000}, and photonic nanostructures
like quantum dots have been investigated.

However, periodically-poled crystals are distinguished from other
photon-pair sources for two reasons
\cite{harris:063602,2007CRPhy8180,2008PhRvL.100r3601N}. High
photon-pair fluxes are reached because of the use of materials
with high nonlinearities due to quasi-phase-matching. On the other
hand, chirping allows to generate photon pairs with extremely wide
spectral bandwidths and subsequently extremely sharp temporal
features. Such photon pairs are useful in many experiments, e.g.,
in quantum optical coherence tomography \cite{Carrasco2004} or
determination of tunnelling times \cite{Steinberg1993}. Wide
spectral bandwidths of entangled photon pairs make them
prospective for parallel processing of quantum information.

In theory, description of chirped periodically-poled crystals has
been restricted to simple more-less analytically soluble models up
to now \cite{harris:063602,2007CRPhy8180,2008PhRvL.100r3601N}.
Properties of photon pairs in the transverse plane have not been
addressed in these models. In the paper, we investigate spectral,
temporal as well as spatial properties of the emitted photon pairs
in detail using a complete theory. Special attention has been
devoted to the role of chirping parameter that is found
extraordinarily useful in tailoring properties of photon pairs.
Splitting of entanglement area that is accompanied by spectral
splitting is predicted in this model. Also temperature
dependencies of intensity profiles experimentally observed in
\cite{2008PhRvL.100r3601N} have been addressed using the model.

The model of spontaneous parametric down-conversion is introduced
in Sec.~II. Quantities characterizing photon pairs are defined in
Sec.~III. Spectral and temporal properties as well as entanglement
of photons in a pair are discussed in Sec.~IV. Properties of
photon pairs in the transverse plane including entanglement area
and temperature dependencies are analyzed in Sec.~V. Conclusions
are drawn in Sec. VI.

\section{Model of spontaneous parametric down-conversion}

We use multi-mode description of spontaneous parametric
down-conversion \cite{mandel}.  Each field, i.e. pump, signal and
idler, is represented by plane-waves with given polarizations.
Nonlinear process is described by first-order perturbation
solution of  Sch${\rm {\ddot{o}}}$dinger's equation in the interaction picture.
Interaction Hamiltonian $ \hat{H}_{\rm int} $ is in the form:
\begin{align} 
  \nonumber \hat{H}_{\rm int}\left(t\right) &=S4\pi^{2}\epsilon_{0}\int_{-L}^{0}dz
  \chi^{\left(2\right)}\left(z\right):\mathbf{E}_{p}^{\left(+\right)}\left(z,t\right)\\
 &\times\hat{\mathbf{E}}_{s}^{(-)}\left(z,t\right)\hat{\mathbf{E}}_{i}^{(-)}\left(z,t\right)+{\rm
  H.c.}.
\label{Eq:1}
\end{align}
Symbol $S$ denotes transverse area of the pump beam,
$\epsilon_{0}$ is permittivity of vacuum, and
$\chi^{\left(2\right)}\left(z\right)$ represents a third-order
tensor of second-order nonlinear susceptibility that is assumed
wavelength-independent. Symbol $ : $ is a shorthand for tensor $
\chi^{(2)} $ with respect to its three indices and H.c. replaces a
Hermitian conjugated term. Symbol $L$ denotes length of the
nonlinear crystal. The strong pump field is described classically
whereas the weak signal and idler fields are treated quantally
using operator amplitudes. An efficient nonlinear interaction is
observed provided that $ x $, $ y $, and $ z $ components of the
wave vectors fulfil at least approximately phase matching
conditions in the corresponding directions. Assuming plane-wave
pumping and sufficiently wide crystal perfect phase matching in
the transverse plane (spanned by $ x $ and $ y $ variables)
occurs.

Positive frequency part of pump-field amplitude $
\mathbf{E}_{P}^{\left(+\right)}$ has the form:
\begin{align}     
 \nonumber
 \mathbf{E}_{p}^{(+)}\left(z,t\right)&=\sum_{\gamma=TM,TE}\int_{-\infty}^{\infty}d\omega_{p}
 \, \tilde{\mathbf{E}}_{p,\gamma}^{(+)}\left(z,\omega_{p}\right)\\ &\times
 \exp{\left[i\left(\left[\mathbf{k}_{p,\gamma}\right]_{z}z-\omega_{p}t\right)\right]}.
\label{Eq:2}
\end{align}
The abbreviation TM (TE) stands for transversal magnetic (electric)
waves, $\omega_{p}$ is pump-field angular frequency,
$\mathbf{k}_{p,\gamma}$ the corresponding wave-vector, and
$\tilde{\mathbf{E}}_{p,\gamma}^{(+)}\left(z,\omega_{p}\right)$ spectrum of
the positive-frequency part of the pump-field amplitude. Strong
coherent pumping is assumed and so the depletion of the pump field can be neglected.

The signal and idler fields are described by negative-frequency
parts of electric-field amplitude operators
$\hat{\mathbf{E}}_{s}^{\left(-\right)}$ and $\hat{\mathbf{E}}_{i}^{\left(-\right)}$,
respectively, that can be written as follows \cite{Vogel}:
\begin{align}   
 \nonumber \hat{\mathbf{E}}_{j}^{(-)}\left(z,t\right)&=\sum_{\xi=TM,TE}
  \int_{-\infty}^{\infty}d\omega_{j}\,
  \sqrt{\frac{\hbar\omega_{j}}{4\pi\epsilon_{0}cn_{\xi}(\omega_{j})S_{\bot}}}\mathbf{e}_{j,\xi}\left(\omega_{j}\right) \\
 &\times \hat{a}_{j,\xi}^{\dagger}\left(\omega_{j}\right)
  \exp{\left[-i\left(\left[\mathbf{k}_{j,\xi}\right]_{z}z-\omega_{j}t\right)\right]},
\label{Eq:3}
\end{align}
$j=s,i$. Symbol $\mathbf{e}_{j,\xi}\left(\omega_{j}\right)$ stands for
polarization vector of a mode with frequency $\omega_{j}$ and
polarization $\xi$ whereas
$\hat{a}_{j,\xi}^{\dagger}\left(\omega_{j}\right)$ represents the
creation operator of this mode. Symbol $\hbar$ denotes reduced
Planck's constant, $n_{\xi}(\omega_{j})$ is the index of
refraction and $ c $ is speed of light.

First-order solution of Sch${\rm {\ddot{o}}}$dinger's equation
gives the wave-function of an entangled two-photon state in the
output plane of the crystal:
\begin{equation}   
 \left|\psi^{(2)}\right\rangle
  =-\frac{i}{\hbar}\int_{-\infty}^{\infty}dt\acute{}\hat{H}_{\rm int}\left(t\acute{}\right)
  \left|\psi_{00}\right\rangle,
\label{Eq:4}
\end{equation}
where $\left|\psi_{00}\right\rangle $ denotes the initial signal-
and idler-field vacuum state. Contributions described by higher
orders of the perturbation solution are assumed to be small due to
relatively low pumping intensity.
Taking into account Eqs. (\ref{Eq:1}), (\ref{Eq:2}), (\ref{Eq:3}), and
(\ref{Eq:4}) the wave function $ \left|\psi^{(2)}\right\rangle $
can be obtained in the form:
\begin{align}   
 \nonumber \left|\psi^{(2)}\right\rangle &=
  \sum_{\alpha,\beta=TM,TE}\int_{-\infty}^{\infty}d\omega_{s}\int_{-\infty}^{\infty}d\omega_{i}\\
 &\times\tilde{\cal A}_{\alpha\beta}\left(\omega_{s},\omega_{i}\right)\hat{a}_{s,\alpha}^{\dagger}
  \left(\omega_{s}\right)\hat{a}_{i,\beta}^{\dagger}\left(\omega_{i}\right)
  \left|\psi_{00}\right\rangle .
\label{Eq:5}
\end{align}
Here, $\tilde{\cal
A}_{\alpha\beta}\left(\omega_{s},\omega_{i}\right)$ denotes a
two-photon spectral amplitude giving probability amplitude of
emitting a signal photon at frequency $\omega_{s}$ and
polarization $\alpha$ together with an idler photon at frequency
$\omega_{i}$ and polarization $\beta$. Two-photon amplitude
$\tilde{ \cal A}_{\alpha\beta} $ can be expressed as
\cite{PhysRevA.56.1534,PhysRevA.59.2359,PhysRevA.56.R21}
:
\begin{align}   
 \nonumber \tilde{\cal A}_{\alpha\beta}\left(\omega_{s},\omega_{i}\right) &=
C_{\alpha,\beta}^{\psi}\left(\omega_{s},\omega_{i}\right)
  \sum_{\gamma=TM,TE}\int_{-\infty}^{\infty}d\omega_{p}\int_{-L}^{0}dz\\
 \nonumber &\times  \int_{-\infty}^{\infty}dk_{n}
  \tilde{\chi}^{\left(2\right)}\left(k_{n}\right):\mathbf{e}_{s,\alpha} \left(\omega_{s}\right)\mathbf{e}_{i,\beta}\left(\omega_{i}\right)\\
   \nonumber&\times\exp{\left[i\left(\left[\mathbf{k}_{p,\gamma}\right]_{z}-\left[\mathbf{k}_{s,\alpha}
  \right]_{z}-\left[\mathbf{k}_{i,\beta}\right]_{z}-k_{n}\right)z\right]}\\
&\times \tilde{\mathbf{E}}_{p,\gamma}^{\left(+\right)}\left(0,\omega_{p}\right)\delta(\omega_{p}-\omega_{s}-\omega_{i})
\label{Eq:6}
\end{align}
where $\tilde{\chi}^{(2)}$ is Fourier
transform of function $\chi^{(2)}(z)$ and $k_{n}$ denotes a grid vector of poling. Alternatively, the crystal
length can be given indirectly by number $ N_L $ of layers.
Normalization function
$C_{\alpha,\beta}^{\psi}\left(\omega_{s},\omega_{i}\right)$ has
the form:
\begin{equation}   
 C_{\alpha,\beta}^{\psi}\left(\omega_{s},\omega_{i}\right)=
  \frac{\pi\sqrt{\omega_{s}\omega_{i}}}{ic\sqrt{n_{\alpha}(\omega_{s})n_{\beta}(\omega_{i})}}.
\label{Eq:7}
\end{equation}

Variations of second-order nonlinear susceptibility $ \chi^{(2)} $
along the $ z $-axis can be described using the formula:
\begin{equation}  
 \chi^{\left(2\right)}\left(z\right)=\chi_{0}^{\left(2\right)}\rm{sign}[\cos\left(k_{n0}z-\zeta
 z^{2}\right)] \rm{rect}\left[\frac{z}{-L}\right].
\label{Eq:8}
\end{equation}
Symbol $k_{n0}$ denotes a basic grid vector of poling and $ k_{n0}
= \pi/l_{0} $ where $ l_{0} $ represents the length of the basic
layer. Parameter $ \zeta $ characterizes chirping of the poled
structure. Function $ {\rm rect}(x) $ equals 1 for $ 0\ge x\ge 1 $
and is zero otherwise. We note that periodical poling can be
reached by applying strong static electric fields to domains
(defined by contacts prepared by lithography) that invert the sign
of nonlinear susceptibility $ \chi^{(2)} $ \cite{2007CRPhy8180}.
 An other method is
based on mapping an optical-field interference pattern to $
\chi^{(2)} $ nonlinearity that can be done provided that a
nonlinear material is placed into a strong uv radiation
\cite{sones:072905}.

\subsection{Simplified description using uniform poling}

Formation of the spectral shape by interference from all layers
can be studied analytically for uniform poling and cw pumping
provided that we restrict ourselves to a scalar modes and consider
propagation along the $ z $ axis only.  The two-photon amplitude $
\tilde {\cal A} $ can then be expressed as:
\begin{align}   
 \nonumber \tilde{\cal A}(\omega_{s},\omega_{i})&=C^{\psi}(\omega_{s},\omega_{i})
 \tilde{E}^{(+)}_{p}(\omega_{p0}) \delta(\omega_{p0}-\omega_{s}-\omega_{i}) \\
&\times\int_{-L}^{0} dz \chi^{(2)}(z)\exp{[i\Delta k z]};
\label{Eq:9}
\end{align}
$\Delta k = k_{p}-k_{s}-k_{i} $. Integration along the crystal in
Eq.~(\ref{Eq:9}) can be replaced by the sum of contributions coming
from each layer and then the two-photon amplitude $ \tilde {\cal A} $
takes the form:
\begin{align} 
 \nonumber \tilde{ \cal A}(\omega_{s},\omega_{i})&=C^{\psi}(\omega_{s},\omega_{i})
  \tilde{E}^{(+)}_{p}(\omega_{p0})\delta(\omega_{p0}-\omega_{s}-\omega_{i}) \\
 \nonumber &\times l_{o}{\rm sinc}\left[\frac{\Delta k l_{o}}{2}\right]
  \frac{\sin\left[(\Delta k l_{o}-\pi)N_L/2\right]}
  {\sin\left[(\Delta k l_{o}-\pi)/2\right]} \\
 &\times \exp{\left[-i(\Delta k l_{0}+\pi)\frac{N_L}{2}-i\frac{\pi}{2}\right]}.
\label{Eq:10}
\end{align}
Symbol $ l_o $ introduced in Eq.~(\ref{Eq:10}) denotes the length of
one layer ($ l_o = L/N_L $). Function $ {\rm sinc }$ in
Eq.~(\ref{Eq:10}) originates in phase mismatch of one typical layer
whereas superposition of down-converted fields from all layers
gives the remaining part of the formula in Eq.~(\ref{Eq:10}) that
resembles the well-known expression describing interference of a
finite number of waves. Maximum value of the two-photon amplitude
$ \tilde{\cal A} $ in Eq.~(\ref{Eq:10}) is reached if the sin
function in the denominator is zero. This happens under the usual
quasi-phase-matching condition $ \Delta k = \pi(n+1)/l_o $, $ n $
being an integer. The number of emitted photon-pairs at the
corresponding frequencies is then proportional to $ N_L^2 $. The
expression in Eq.~(\ref{Eq:10}) also indicates that the spectral
width of the signal or idler field is proportional to $ 1/N_L $
for cw pumping.

\section{Characteristics of photon pairs emitted in periodically poled crystals}

Number of generated photon pairs $N$ (per 1~s in cw regime) can be
easily calculated from two-photon spectral amplitude $\tilde{\cal
A} $ given in Eq.~(\ref{Eq:6}):
\begin{equation}   
 N=\int_{-\infty}^{\infty}d\omega_{s}\int_{-\infty}^{\infty}d\omega_{i}|
 \tilde{\cal A}\left(\omega_{s},\omega_{i}\right)|^{2}.
\label{Eq:11}
\end{equation}
Energy spectrum $S_{s,i}\left(\omega_{s,i}\right)$ for the
signal or idler field is determined similarly with the formula:
\begin{equation}   
 S_{s,i}\left(\omega_{s,i}\right)=\hbar\omega_{s,i} \int_{-\infty}^{\infty}d\omega_{i,s}
 |\tilde{\cal A}\left(\omega_{s},\omega_{i}\right)|^{2}.
\label{Eq:12}
\end{equation}

Schmidt's decomposition of the two-photon spectral amplitude
$\tilde{\cal A} $ into the dual basis can be used to
quantify entanglement between signal and idler photons. Two-photon
spectral amplitude $\tilde{\cal A}(\omega_{s},\omega_{i})$ can then be
expressed as
\begin{equation}   
 \tilde{\cal A}(\omega_{s},\omega_{i})={\displaystyle
 \sum_{n=1}^{\infty}} \lambda_{n}\psi_{n}\left(\omega_{s}\right)
 \phi_{n}\left(\omega_{i}\right) ,
\label{Eq:13}
\end{equation}
where $\lambda_{n}$ are eigenvalues corresponding to base
functions $\psi_{n}$ and $\phi_{n}$. The base functions $\psi_{n}$
and $\phi_{n}$ are determined as solutions of the integral
equations:
\begin{align}   
 \int K_{s}\left(\omega,\acute{\omega}\right)\psi_{n}\left(\acute{\omega}\right)
  d\acute{\omega}&=\lambda_{n}^{2}\psi_{n}\left(\omega\right),
  \label{Eq:14}\\
 \nonumber \\
 \int K_{i}\left(\omega,\acute{\omega}\right)\phi_{n}
  \left(\acute{\omega}\right)d\acute{\omega}&=\lambda_{n}^{2}\phi_{n}\left(\omega\right).
  \label{Eq:15}
\end{align}
Entropy of entanglement $ E $ \cite{PhysRevLett.84.5304} defined
as:
\begin{equation}   
 E=-{\displaystyle \sum_{n=1}^{\infty}}\lambda_{n}^{2}\log_{2}\lambda_{n}^{2}
\label{Eq:16}
\end{equation}
is used for quantification of entanglement. The higher the entropy
$ E $, the stronger the entanglement. The number of effective
independent modes useful in Schmidt's decomposition is given by
cooperation parameter $ K $ \cite{law:127903}:
\begin{equation}  
 K={\displaystyle \frac{1}{\sum_{n=1}^{\infty}\lambda_{n}^{4}}}.
\label{Eq:17}
\end{equation}

Temporal properties of photons in a pair can only be measured
indirectly using interferometers. Relative time delay between
photons in a pair can be determined in Hong-Ou-Mandel
interferometer \cite{hong}. Normalized coincidence-count rate $
R_{n}\left(\tau\right)$ in this interferometer is expressed in the
form:
\begin{equation}   
 R_{n}\left(\tau\right)=1-\rho\left(\tau\right)
\label{Eq:18}
\end{equation}
and
\begin{align}   
 \nonumber \rho\left(\tau\right)&=\frac{1}{R_{0}}\int_{-\infty}^{\infty}dt_{a}
  \int_{-\infty}^{\infty}dt_{b} \\
 \nonumber &\times Re\left[{\cal A}\left(t_{a},t_{b}-\tau\right){\cal A}^{*}\left(t_{b},t_{a}-\tau\right)\right] \\
 \nonumber &=\frac{1}{\left(2\pi\right)^{2}R_{0}}\int_{-\infty}^{\infty}d\omega_{s}
  \int_{-\infty}^{\infty}d\omega_{i}\\
 &\times{\rm Re}\left[\tilde{\cal A}\left(\omega_{s},\omega_{i}\right)
  \tilde{\cal A}^{*}\left(\omega_{i},\omega_{s}\right)\exp(i\tau\left[\omega_{i}-\omega_{s}\right])
  \right];
\label{Eq:19}
\end{align}
Re denotes the real part of an argument. Symbol $R_{0}$ is a
normalization constant; $ R_0 = \lim_{\tau\rightarrow\infty}
\rho(\tau) $.

\section{Temporal and spectral properties of photon pairs,
entanglement}

The analysis is done for a periodically-poled nonlinear uniaxial
crystal LiNbO$ {}_{3}$. Having in mind optical-fiber applications
we consider cw plain-wave pumping at central wavelength
$\lambda_{p0}=752.5$~nm. The signal and idler fields have
wavelengths $\lambda_{s0}=\lambda_{i0}=1505$~nm
($\lambda_{s0}=1392.1$~nm and $\lambda_{i0}=1637.8$~nm) in the
spectrally-degenerate (spectrally-non-degenerate) case. Assuming
uniform poling the length identical for all layers is determined
such that quasi-phase matching is reached for TM polarizations of
all interacting fields at given central frequencies assuming
collinear propagation. Since the studied material is anisotropic
and we consider the optical axis parallel to the $ y $ axis,
polarizations of all interacting fields correspond to
extra-ordinary waves in collinear geometry. The length of layers
for the spectrally-degenerate (spectrally-non-degenerate) case
equals 8.9000 $\rm{\mu m}$ (8.9286 $\rm{ \mu m }$).

If chirping in poling is assumed, the length of layer in the
middle of the structure equals that for uniform poling. The value
of used chirping parameter $ \zeta $ ($\zeta=1\times 10^{-6} $~$
\mu {\rm m}^{-2}$) corresponds to the case in which shorter layers
occur at the beginning of the crystal whereas longer layers are at
the end of the crystal . We note that if the value of chirping
parameter $ \zeta $ is sufficiently small, layer lengths depend
more-less linearly on their positions. We also note that the
effect of multiple reflections at crystal boundaries that leads to
fast spectral oscillations (typical for a Fabry-Perot etalon
\cite{Saleh}) occurs in periodically-poled crystals. However, we
suppress them in our analysis by making local averaging of the
signal- and idler-field spectra. This is in accord with usual
experimental conditions in which these oscillations are smoothed
out by a final spectral resolution of detectors.

Note that, the Fourier transform of Eq.~(\ref{Eq:8}) contains
several spectral components. The component $+k_{n}$ assures
phase-matching at the central frequencies mentioned above.
However, the component with $-k_{n}$ also exists but cannot
provide phase matching in the considered frequency range for the
studied nonlinear crystal and so we omit it in further
investigations.

\subsection{Spectrum and number of generated photon pairs}

\begin{figure}  
\begin{center}
a)%
\includegraphics[clip,scale=0.75]{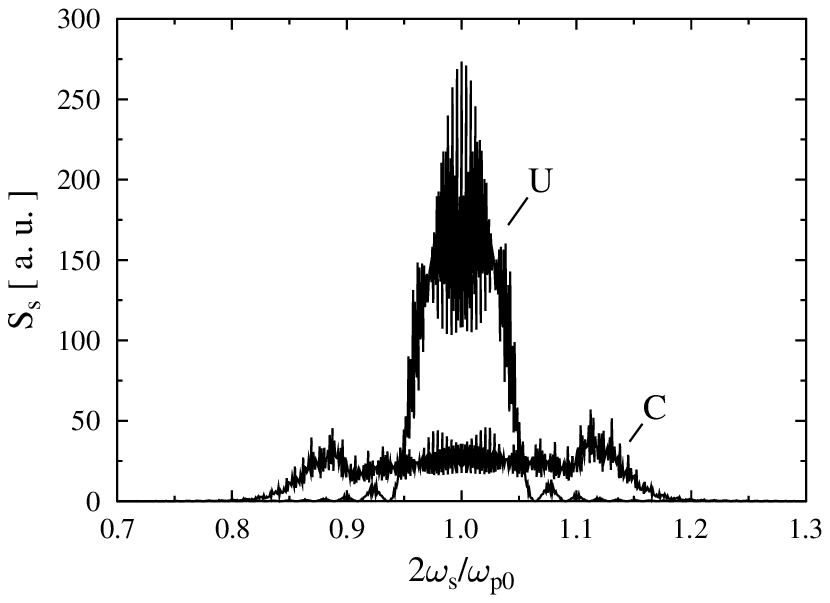}
b)%
\includegraphics[clip,scale=0.75]{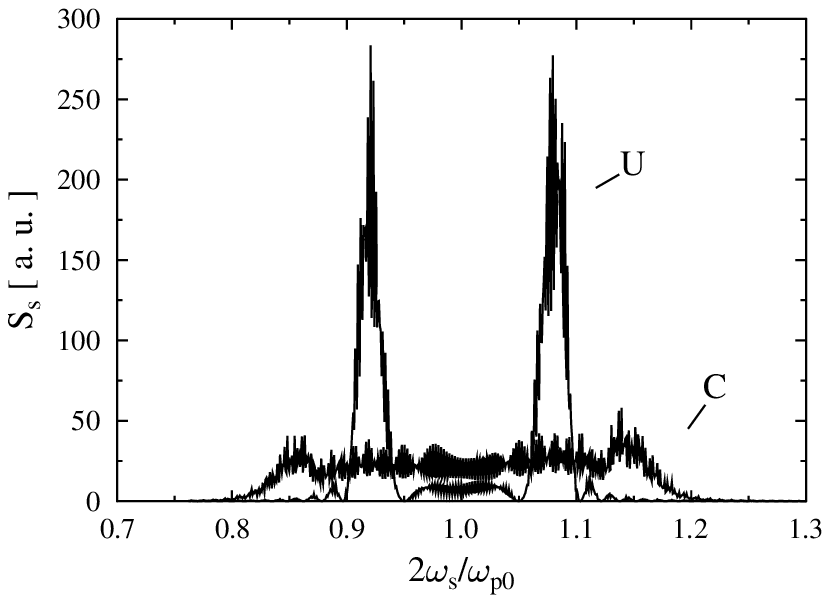}
c)%
\includegraphics[clip,scale=0.75]{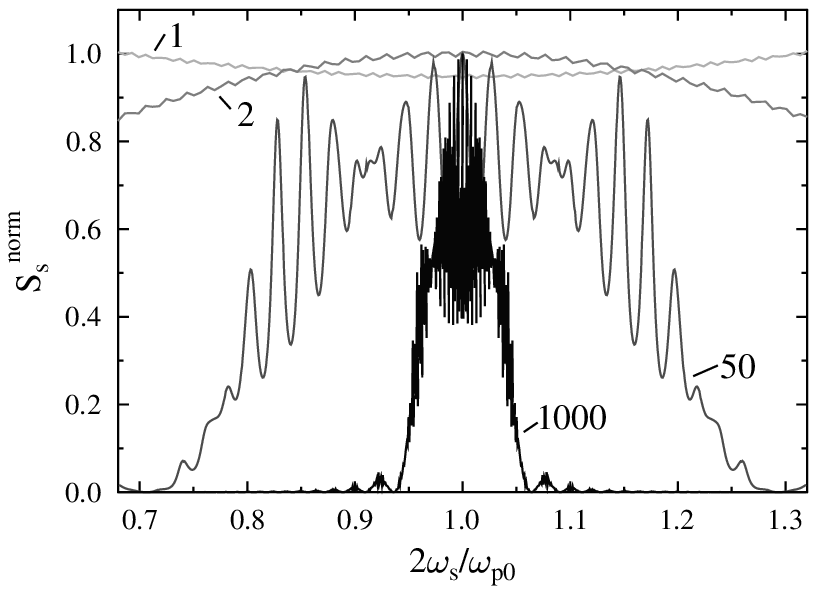}

\end{center}

 \caption{Energy spectrum $ S_s $ for a) spectrally
 degenerate and b) spectrally non-degenerate down-conversion.
 Curves denoted as U (C) are appropriate for uniform (chirped)
 poling of the crystal; $ N_L = 1000 $ in a) and b).
 Forming of normalized energy spectrum $ S_s^{\rm norm} $ with the increasing number
 $ N_L $ of layers is documented
 in c) for uniform poling, spectral degeneration, and $ N_{L} = 1,$ 2, 50, and 1000.}
 \label{fig:1}
\end{figure}

Typical spectra for both frequency-degenerate and
frequency-non-degenerate down-conversion for uniform as well as
chirped poling are shown in Fig.~\ref{fig:1}.

Fast modulation in these spectra has its origin in interference of
two-photon amplitudes coming from different nonlinear layers of
the crystal as expressions in Eqs.~(\ref{Eq:10}) and (\ref{Eq:12})
indicate. In frequency-non-degenerate case and for uniform poling,
there occur two peaks centered at different frequencies (see
Fig.~\ref{fig:1}b). These contributions mutually interfere in the
frequency region where they overlap. Chirped poling leads to wide
broadening of the signal- and idler-field spectra, as demonstrated
in Fig.~\ref{fig:1}a,b. The reason is that two-photon amplitudes
from different layers have shifted ranges of the emitted signal-
and idler-field frequencies and so when two-photon amplitudes from
all layers are superposed, a much broader range of frequencies is
covered. However this broadening is at the expense of lowering
photon-pair emission rates. This occurs because nonlinear layers
of different lengths and generating different two-photon
amplitudes do not admit perfect constructive interference at
optimum frequencies (as it occurs for uniform chirping). We also
note that wider spatial emission angles occur for the higher
values of chirping and so a considerable portion of the emitted
signal does not contribute to the studied collinear case.

Successive forming of the signal-field spectrum as the number $
N_L $ of layers increases is depicted in Fig.~\ref{fig:1}c for
uniform poling. Already two layers (that form the basic element of
the structure) provide a spectrum that prefers the area around the
central frequency. As the curves in Fig.~\ref{fig:1}c document the
larger the number $ N_L $ of layers the narrower the spectrum
[compare the expression in Eq.~(\ref{Eq:10})].

 \begin{figure}  
 a)%
\includegraphics[scale=0.75]{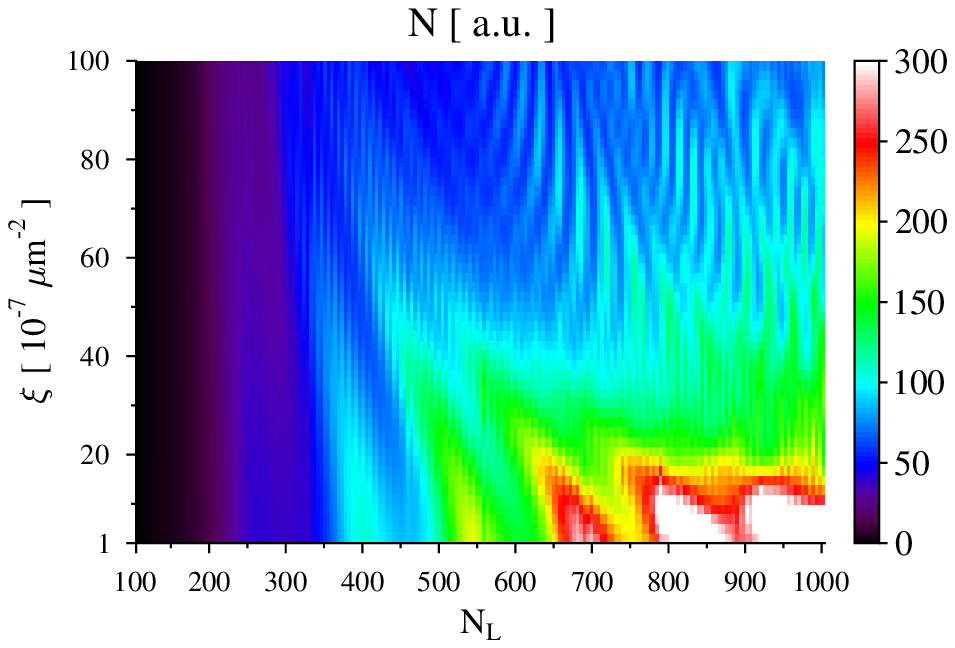}
b)%
\includegraphics[scale=0.75]{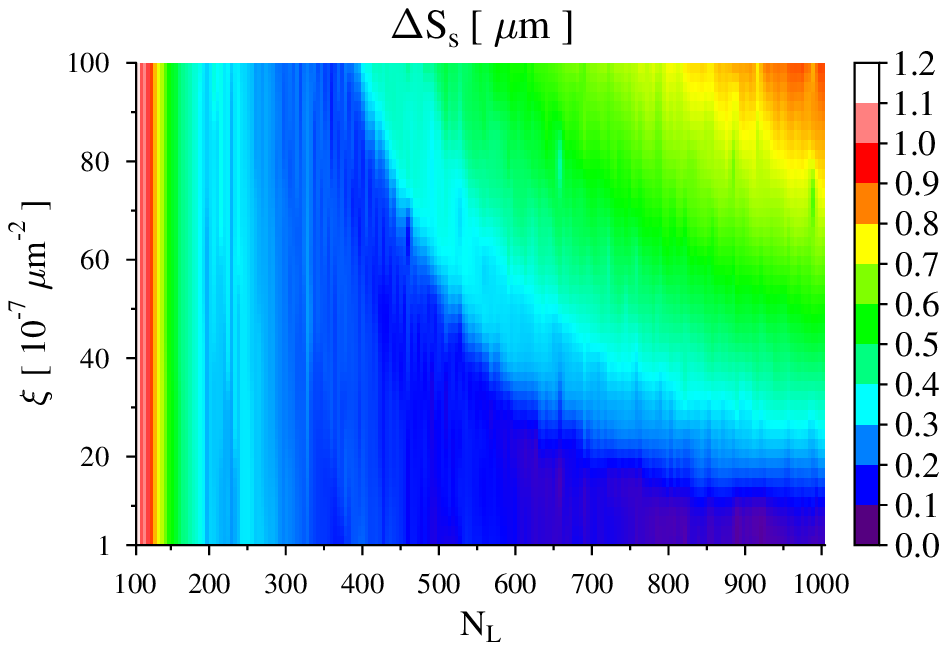}
 \caption{(Color online) Contour plots of a) number $ N $ of generated photon pairs and b) spectral width
 $ \Delta S_{s} $ (FWHM) of the signal field as functions of the number of layers $N_L$
 and chirping parameter $\zeta$. Frequency degenerate case is assumed.}
\label{fig:2}
\end{figure}

Number $ N $ of generated photon pairs and signal- and idler-field
spectral widths are the most important parameters of a photon-pair
source in many applications. We show their dependence on the
number $ N_L $ of layers (giving length $ L $ of the crystal) and
chirping parameter $\zeta$ in Fig.~\ref{fig:2}. We can observe the
following general tendencies. Provided that $ \zeta $ is fixed
increasing number $ N_L $ of layers gives higher numbers $ N $ of
generated photon pairs but lowers spectral widths of the signal
and idler fields. On the other hand and keeping $ N_L $ fixed
increasing values of chirping parameter $ \zeta $ lead to
decreasing of the number $ N $ of generated photon pairs and
broader signal- and idler-field spectra.

Defined spectral widths of the signal and idler fields are
required in many applications. The spectral widths are determined
by the lengths of the first and last nonlinear layers in a crystal
with chirped poling. Number $ N $ of generated photon pairs
increases naturally with increasing number $ N_L $ of layers.
Since lengths of the first and last layers are fixed, a larger
number $ N_L $ of layers means a lower value of chirping parameter
$ \zeta $. Our numerical investigations have revealed that the
product of number $ N $ of generated photon pairs and chirping
parameter $ \zeta $ is roughly constant ($ N\zeta \approx {\rm
const} $) in this case. These observations allow to design easily
structures with defined spectra and photon-pair emission rates.

\subsection{Fourth-order interference of a photon pair}

Hong-Ou-Mandel interferometer giving fourth-order interference
pattern allows to determine entanglement time from the profile of
fourth-order interference pattern obtained by varying a mutual
time delay between the signal and idler photons. Entanglement time
gives a typical time window in which an idler photon is detected
in this specific interferometer provided that its twin signal
photon has already been detected. Entanglement time does not bear
any information about the instant of a signal-photon detection.
Entanglement time defined this way does depend on dispersion
present in the paths of the signal and idler photons in general.
However, if the same amount of dispersion is present in the
signal-photon and idler-photon paths entanglement time is not
affected \cite{PhysRevA.59.2359}. In case of cw pumping,
entanglement time is even not sensitive to even orders of material
dispersion in the paths of two photons from a pair
\cite{PhysRevA.45.6659}. On the other hand, also sum-frequency
generation can be used to define another typical time constant
called correlation time (defined in optical-field coherence
theory) \cite{Harris:07}. Correlation time characterizes a true
temporal shape of a biphoton (as given by the amplitude of a
two-photon temporal amplitude) and is sensitive to dispersion.
Careful dispersion compensation is experimentally required to
observe the sharpest temporal features as possible.

\begin{figure}   
a)%
\includegraphics[scale=0.75]{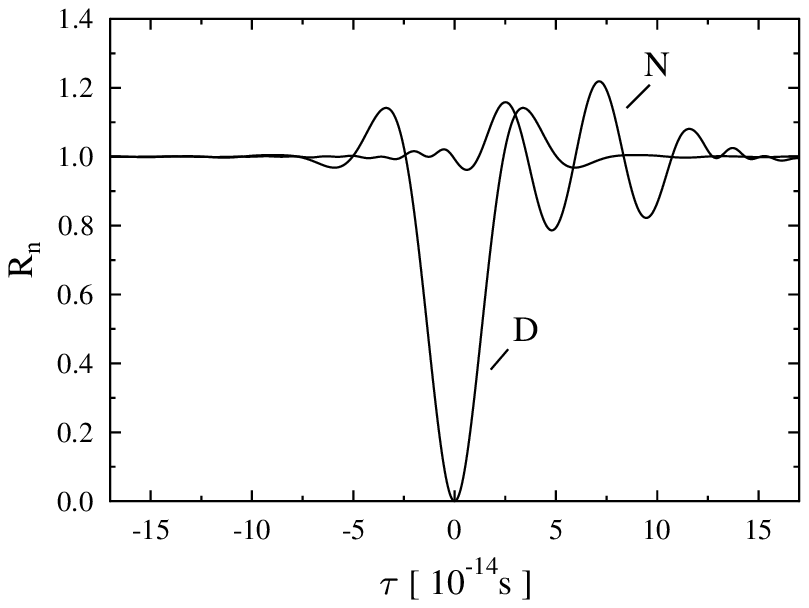}
b)%
\includegraphics[scale=0.75]{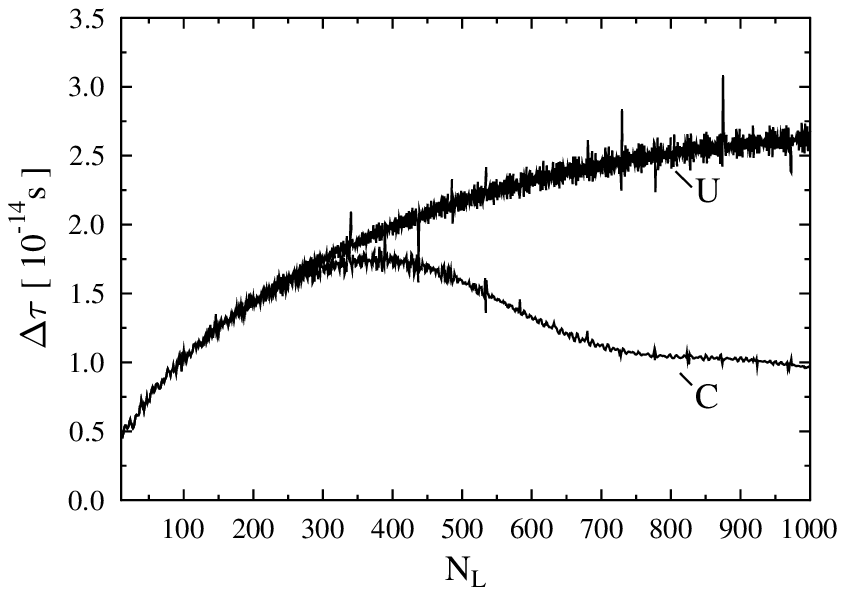}
c)%
\includegraphics[scale=0.75]{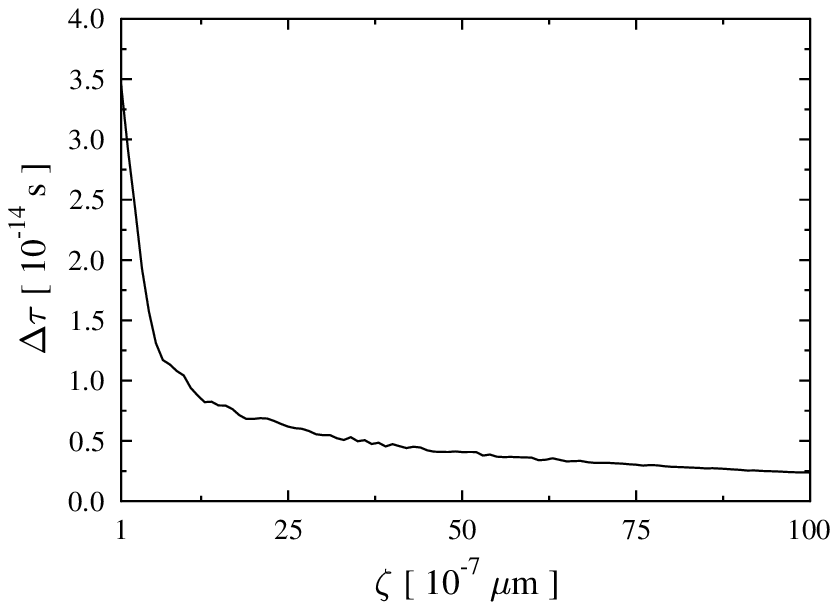}

 \caption{a) Normalized coincidence-count rate $ R_n $ in Hong-Ou-Mandel interferometer
 as a function of relative time delay $ \tau $ for spectrally degenerate
 (D) and non-degenerate (N) down-conversion in a uniformly poled
 crystal. b) Width $ \Delta\tau $ (FWHM) of the interference dip
 in normalized coincidence-count rate $ R_n $ as a function of number $ N_L $ of layers
 for uniform (U) and chirped (C) poling. c) Width $ \Delta\tau $ as it depends on chirping
 parameter $ \zeta $. Frequency degeneration is assumed; $ N_L = 1000
 $, $\zeta = 1\times 10^{-6}$ $\mu m^{-2}$}.
\label{fig:3}
\end{figure}

If we consider frequency degenerate case, both photons in a pair
are indistinguishable (are described by identical wave-functions)
and we observe visibility 100$\%$ in a Hong-Ou-Mandel
interferometer. Different central frequencies of the signal and
idler fields lead to partial distinguishability of two photons
and, as a consequence, the loss of visibility occurs (see
Fig.~\ref{fig:3}a). Moreover, the signal and idler photons
propagate at different group velocities and occur at the output of
the crystal with nonzero relative mutual delay, that is determined
from the shift of the dip position in the interference pattern.
Also oscillations at the difference of the central signal- and
idler-field frequencies occur. Low visibility is usually observed
in this case because of complex interference of two-photon
amplitudes coming from different layers. We note that the
inclusion of third- and higher-odd-order dispersion terms has lead
to small oscillations visible in Fig.~\ref{fig:3}a. Width of the
coincidence-count interference pattern $ R_n(\tau) $ is mainly
given by first-order dispersion properties of the crystal (for
details see, e.g., \cite{PhysRevA.56.1534})

Width $ \Delta\tau $ of the coincidence-count rate $ R_n $
increases with the increasing number $ N_L $ of layers concerning
uniform poling (see Fig.~\ref{fig:3}b), in agreement with
decreasing signal- and idler-field spectral widths. On the other
hand, width $ \Delta\tau $ increases for smaller numbers $ N_L $
of layers whereas it decreases for larger numbers $ N_L $ of
layers for chirped poling. This behavior can be related to the
behavior of the signal- and idler-field spectral widths (see
Fig.~\ref{fig:2}b): the wider the spectrum the smaller the width $
\Delta\tau $. If number $N_{L}$ of layers
 is sufficiently large, the width $\Delta \tau$ decreases with increasing values
of chirping parameter $\zeta$ that makes spectral widths wider
(see Fig.~\ref{fig:3}c). Several applications (e.g, quantum
optical coherence tomography \cite{Carrasco2004}, determination of
ultrashort temporal intervals \cite{Steinberg1993}) require as
short entanglement times as possible. Highly chirped crystals are
then needed as sources of photon pairs giving the widest possible
signal- and idler-field spectra.

\subsection{Entropy of entanglement}

Entropy of entanglement $ E $ can be used to quantify the strength
of mutual spectral entanglement between the signal and idler
photons. Technically, integral equations in Eqs.~(\ref{Eq:14}) and
(\ref{Eq:15}) are discretized and converted this way into the
problem of matrix diagonalization. Division has to be sufficiently
close which is indicated by a large number of eigenvalues lying
close to zero. Eigenvalues of Schmidt's decomposition do not
depend on additional phase variations introduced by dispersion
present in paths of the signal and idler photons. This can be
deduced from the definition of kernels $ K_s $ and $ K_i $
\cite{jr.:013803} that are proportional to the reduced statistical
operators of the signal and idler photons, respectively. These
phase variations only affect phases of the complex base functions
$ \psi_n $ and $ \phi_n $ occurring in the decomposition of
two-photon spectral amplitude $ \tilde{\cal A} $ in
Eq.~(\ref{Eq:13}).

\begin{figure}   
 \includegraphics[scale=0.75]{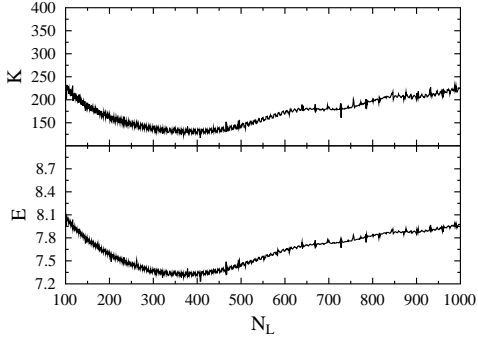}
 \caption{Entropy $ E $ and cooperativity parameter $ K $ as
 they depend on number $ N_L $ of layers.
 Frequency degeneration as well as chirped poling are assumed;
  $\zeta = 1 \times 10^{-6}$~$\mu m^{-2}$.}
\label{fig:4}
\end{figure}

Entropy of entanglement $ E $ behaves similarly as signal- and
idler-field spectral widths. The wider the spectra, the higher the
entropy $ E $. This means, that the higher the number $ N_L $ of
layers in a uniformly poled crystal the lower the entropy $ E $.
Assuming chirped poling, entropy $ E $ decreases until a certain
number $ N_{L0} $ of layers is reached and then increases (see
Fig.~\ref{fig:4}). Cooperativity parameter $ K $ behaves similarly
(see Fig.~\ref{fig:4}) as entropy $ E $. We note that the wider
the signal- and idler-field spectra the larger the number of
eigenvalues $ \lambda_n $ needed in Schmidt's decomposition and as
a consequence greater values of cooperativity parameter $ K $ and
entropy $ E $ occur.

\section{Spatial properties of photon pairs, correlation area}

\subsection{Intensity cross-section of the down-converted fields}

\begin{figure}   
a)%
\includegraphics[scale=0.75]{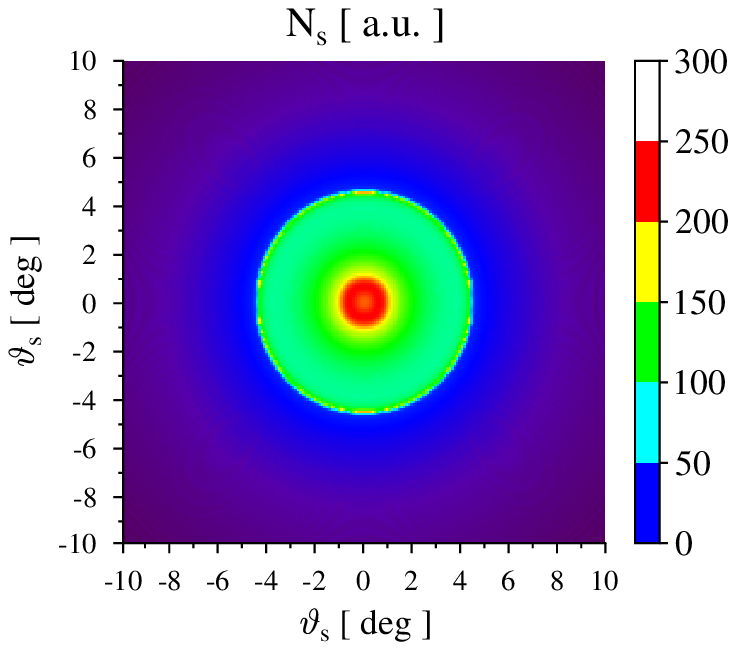}
b)%
\includegraphics[scale=0.75]{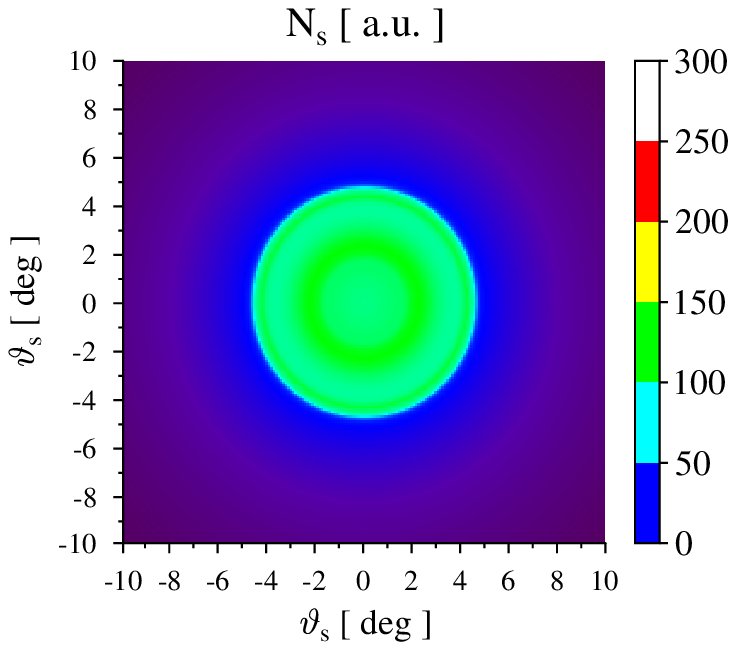}
c)%
\includegraphics[scale=0.75]{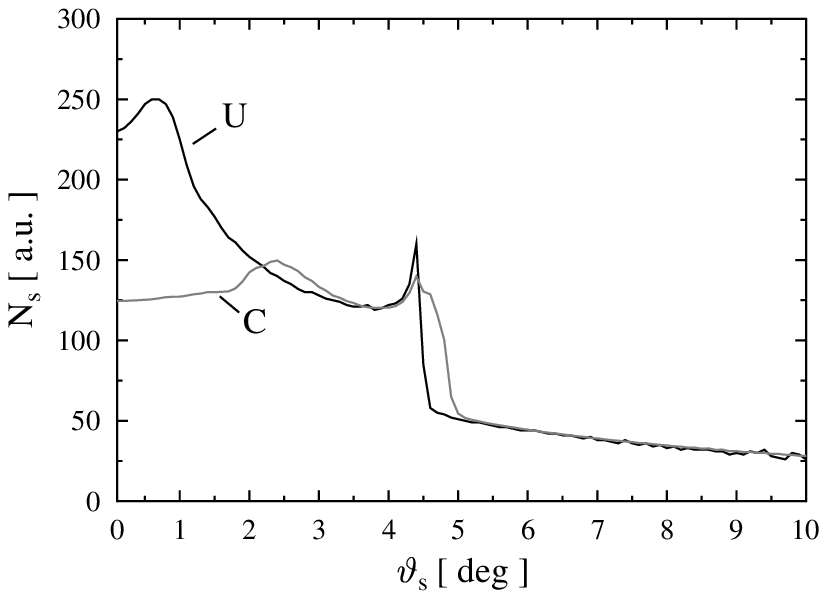}
 \caption{(Color online) Contour plot of the cross-section of the signal-field photon number
 $ N_s $ projected onto hemisphere with radial emission angle $ \vartheta_s $ and azimuthal emission
 angle $ \psi_s $ (determines rotation angle around the center of
 the graph) for a) uniform and b) chirped poling. c) Cut of signal-field photon number
 $ N_s $ along the $ y $ axis ($ \psi_s=0 $~deg) traced by radial emission angle $ \vartheta_s $
 for uniform (U) and chirped (C) poling. Frequency degeneration is assumed;
 $\zeta = 1\times 10^{-6}$~$\mu m^{-2}$, $ N_L = 1000 $.}
\label{fig:5}
\end{figure}

A typical cross section of the signal-field photon number $ N_s $
for uniform poling is displayed in Fig.~\ref{fig:5}a where
p-polarized signal and idler fields degenerate in frequency have
been assumed. This means that the signal and idler photons
propagate as extraordinary waves. The signal and idler fields
propagate in general as a mixture of TE and TM waves at different
emission angles, but the difference between TE and TM waves is
small because of small values of radial emission angles and so the
pattern of cross section is nearly spherically symmetric and
occurs for radial emission angles $ \vartheta_s $ lower than
5~deg. The highest numbers $ N $ of generated photon pairs are
found in the vicinity of $ \vartheta_s = 0 $~deg because the
structure is optimized for collinear interaction.

Comparison of Figs.~\ref{fig:5}a and b reveals that chirping of
poling practically does not change the emission angles. It mainly
reduces the number $ N $ of generated photon pairs in the middle
of the emission pattern, i.e. at the emission angles at which the
conditions for constructive interference of fields from different
layers are worse in case of chirped poling. As a consequence, a
relatively wide plateau replaces a typical peak found for uniform
poling (see Fig.~\ref{fig:5}c for cuts of the cross-sections). We
note that the idler twin of a signal photon is emitted into a
direction given by phase-matching conditions both in the
transverse plane and along the $ z $ axis.

\begin{figure}   
\includegraphics[scale=0.75]{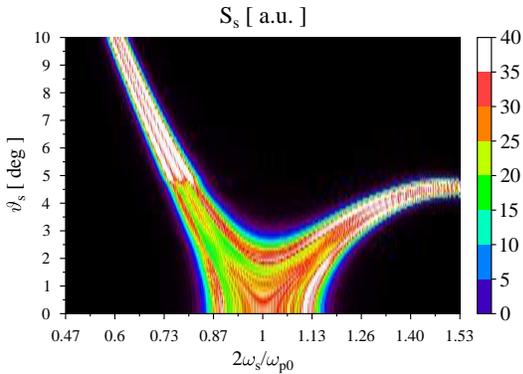}
 \caption{(Color online) Contour plot of signal-field energy spectrum $ S_s $
 as a function of signal-field emission angle $\vartheta_{s}$ assuming chirped poling
 and degenerate spectrum; $\zeta = 1\times 10^{-6}$ $\mu m^{-2}$, $ N_L = 1000 $.}
\label{fig:6}
\end{figure}

Signal- and idler-field emission spectra are broad for all
emission angles as Fig.~\ref{fig:6} showing signal-field energy
spectrum $ S_s $ documents. There occurs splitting of the
signal-field energy spectrum $ S_s $ for radial emission angles $
\vartheta_s $ greater than 2~deg. The use of frequency filters
thus allows to modify the emission-cone cross-section. For
example, removal of the central part of the energy spectrum $ S_s
$ leaves the emission-cone cross section in the form of a
concentric ring.

\subsection{Correlation area of photon pairs}

Correlation area determines possible emission directions of an
idler photon [described by radial ($ \vartheta_i $) and azimuthal
($ \psi_i $) emission angles] which twin, i.e. signal photon, has
been observed in a given direction defined by emission angles $
\vartheta_s $ and $ \psi_s $. Because the crystal is not
infinitely long and also the pump beam is not infinitely broad
entanglement area has finite width that can be characterized by
typical widths of the distribution giving probability of emission
of an idler photon in a given direction. We denote them as $
\Delta\vartheta_i $ and $ \Delta\psi_i $ in the radial and
azimuthal directions, respectively. Qualitatively, the longer the
crystal the smaller the correlation area. Also the broader the
transverse profile of the pump beam the smaller the correlation
area.

\begin{figure}   
a)%
\includegraphics[scale=0.75]{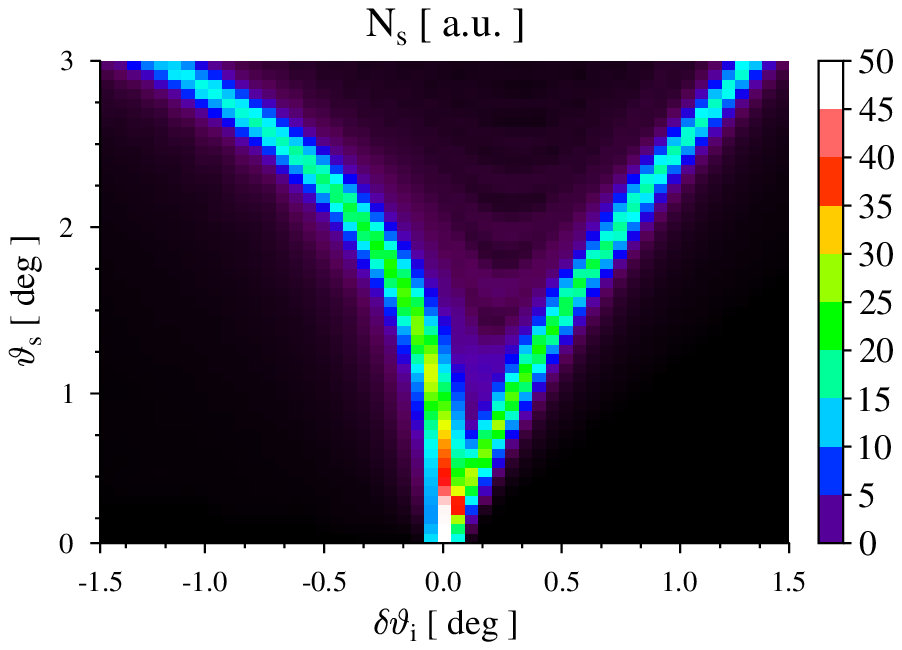}
b)%
\includegraphics[scale=0.75]{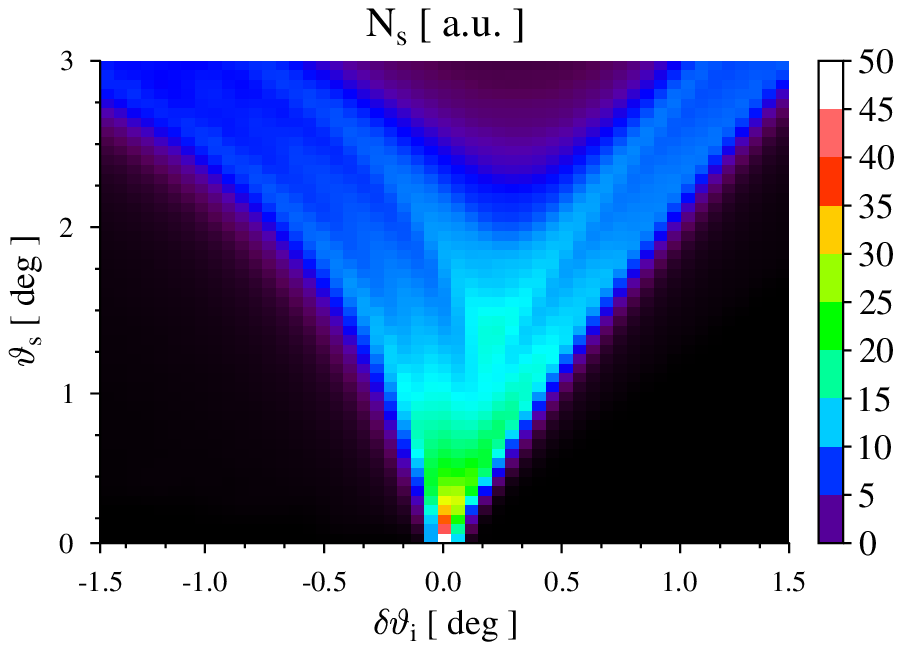}
\caption{(Color online) Contour plots showing the cut of
correlation area along the idler-field radial emission angle $
\vartheta_i $ ($ \vartheta_i = \vartheta_{i}^{\rm opt} +
\delta\vartheta_i $, $ \vartheta_{i}^{\rm opt} $ is given by
optimum phase-matching conditions corresponding to a given $
\vartheta_s $) as it depends on signal-field radial emission angle
$\vartheta_{s} $. Frequency-degenerate emission of p polarized
signal and idler photons along the $ y $ axis is considered both
for a) uniform and b) chirped poling; $N_{L}$ =1000, $\zeta$ =
$1\times 10 ^{-6}$ $\mu m^{-2}$. }
 \label{fig:7}
\end{figure}

\begin{figure}   
\includegraphics[scale=0.75]{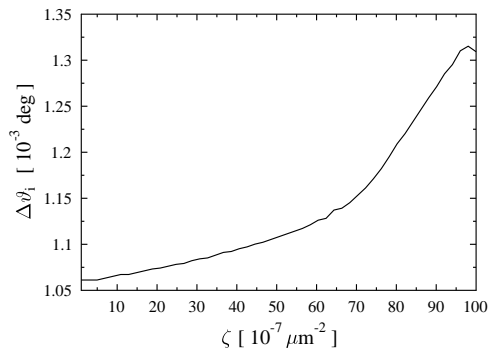}
\caption{Width $\Delta\vartheta_{i}$ of correlation area in radial
direction as a function of $\zeta$ for collinear geometry and
frequency non-degenerate case. The signal and idler photons are p
polarized; $N_{L}$ = 1000.}
 \label{fig:8}
\end{figure}

There occurs splitting of correlation area into two parts in
radial direction for emission angles that are farther from the
optimum emission angle $ \vartheta_{s0} $ for which the structure
was designed. Considering uniform poling and structure optimized
for collinear interaction, this splitting is clearly visible in
Fig.~\ref{fig:7}a for radial emission angles $ \vartheta_s $
greater than 0.5~deg. Splitting in radial emission angle is
accompanied by splitting in frequencies, i.e. frequency spectra of
each part are separated (compare Figs.~\ref{fig:7}b and
\ref{fig:6}). Chirping of poling leads to blurring of correlation
areas, as documented in Fig.~\ref{fig:7}b. This is natural,
because chirping provides broader spectra of the signal and idler
fields that allow to fulfill phase-matching conditions in a wider
area of emission angles. As is evident from Fig.~\ref{fig:7}b
chirping partially conceals splitting of correlation area. Because
width of correlation area is a monotonous function of chirping
parameter $ \zeta $ (see Fig.~\ref{fig:8}), photon pairs with the
required width of correlation area can be generated from poled
crystals with appropriate chirping. We note that width of
correlation area is an important parameter whenever wave-functions
of two photons overlap.

\subsection{Temperature tuning of the intensity cross-section}

\begin{figure}     
a)%
\includegraphics[scale=0.75]{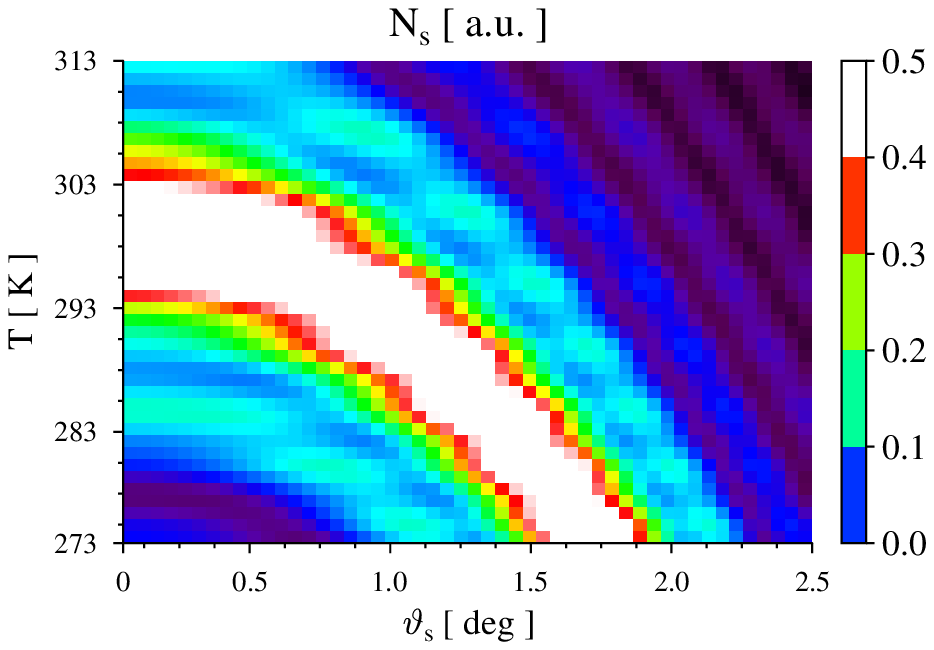}
b)%
\includegraphics[scale=0.75]{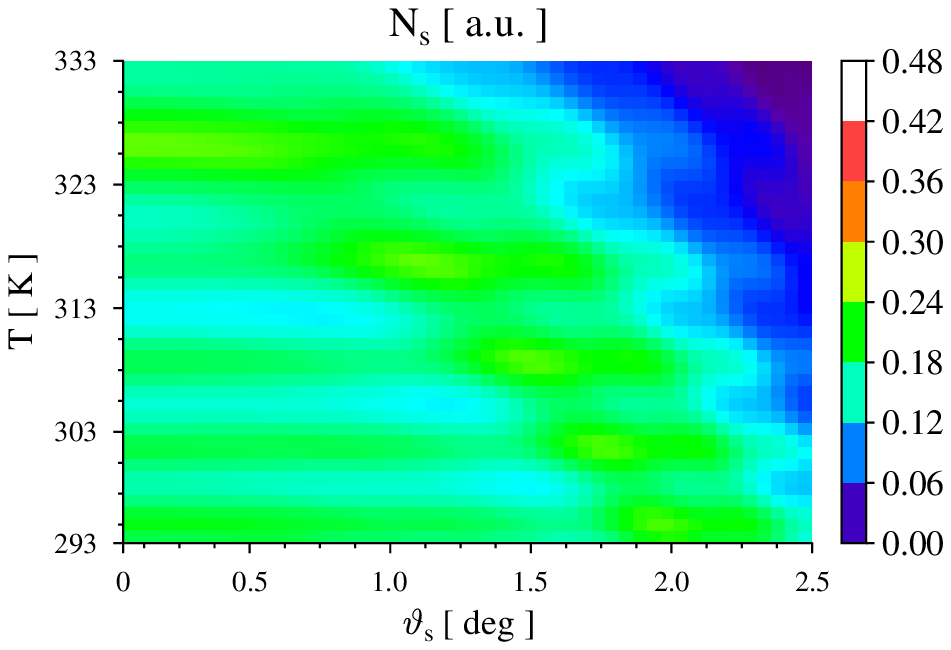}

\caption{(Color online) Contour plots giving the signal-field
photon number $ N_s $ along the $ y $ axis parameterized by radial
emission angle $ \vartheta_s $ as it depends on temperature $ T $.
The signal and idler photons are p polarized. Chirped poling with
a) $\zeta$ = $1\times10^{-7}$ $\rm{\mu m ^{-2}}$ and b) $\zeta$ =
$1\times10^{-6}$ $\rm{\mu m ^{-2}}$ is considered; $N_{L}$ = 1000.
Spectrally degenerate photons are filtered with narrow-band filter
14~THz wide. } \label{fig:9}
\end{figure}

Properties of the emitted photon pairs in the transverse plane can
be controlled by temperature of the crystal as has been
experimentally demonstrated in \cite{2008PhRvL.100r3601N} for
chirped periodically poled stoichiometric LiTaO${}_{3}$. The
change of temperature causes changes in values of indices of
refraction and also widths of layers undergo a certain temperature
shift. As a consequence, temperature can be used to tune a sample
into resonance where phase-matching conditions are perfectly
fulfilled. If temperature changes from its optimum value, shift in
emission directions is observed (see Fig.~\ref{fig:9}). If the
crystal is optimized for collinear geometry, a spot occurring
under optimum conditions is replaced by a concentric ring. The
greater the declination of temperature from its optimum value the
greater the diameter of this ring. If the crystal is highly
chirped, this ring is smoothed and a disc is observed. Width of
the spot as well as width of the ring depend strongly on bandwidth
of frequency filters placed in front of detectors. The wider the
bandwidth, the wider the spot and ring.

\section{Conclusions}

A comprehensive study of physical properties of entangled photon
pairs generated in chirped periodically-poled nonlinear crystals
has been done using a rigorous model of spontaneous parametric
down-conversion. Spectral, temporal, and spatial characteristics
have been addressed. Ultra-wide spectra, high photon fluxes as
well as entanglement areas can be conveniently controlled by
chirping parameter. Ultra-wide spectra are responsible for high
values of entanglement of photons in a pair that can be useful for
quantum-information processing. The experimentally observed
intensity profiles of the down-converted fields in the form of
discs or rings with temperature-dependent parameters have been
theoretically explained in this model. It has been revealed that
complex interference of photon pairs from different layers leads
to splitting of correlation area that is accompanied by spectral
splitting. The presented results have shown that chirped
periodically-poled crystals are not only promising as useful
sources of photon pairs for many applications but they also remain
interesting for investigations of their rich physical properties.

\acknowledgments Support by projects IAA100100713 of GA AV \v{C}R,
COST 09026, MSM6198959213, and LC06007 of the Czech Ministry of
Education is acknowledged. J.P. also thanks project 1M06002 of the
Czech Ministry of Education.

\newpage
\bibliography{Svozilik}
\bibliographystyle{apsrev}

\end{document}